\documentstyle[11pt,newpasp]{article}

\begin{document}

\title{Confirming EIS Clusters. Multi-object Spectroscopy}

\author{A. Biviano, M. Ramella, W. Boschin}
\affil{Osservatorio Astronomico di Trieste,
Italy}
\author{S. Bardelli}
\affil{Osservatorio Astronomico di Bologna,
Italy}
\author{M. Scodeggio, L.N. da Costa,  L.F. Olsen, M. Nonino}
\affil{European Southern Observatory, Garching bei M\"unchen, Germany}
\author{S. Borgani}
\affil{Istituto Nazionale di Fisica Nucleare, Italy}
\author{M. Girardi}
\affil{Dipartimento di Astronomia, Universit\`a degli Studi di Trieste,
 Italy}


\vspace{1cm}

\noindent Clusters of galaxies  arise from exceptionally high peaks of the
primordial fluctuation density field. Their properties as a function
of redshift, $z$, are therefore highly sensitive to the nature of such
cosmic fluctuations. It is therefore very important to have a sample
of galaxy clusters covering as wide a redshift range as possible.

Recently, Olsen et al. (1999) and Scodeggio et al.  (1999) have
identified clusters in 2D from the I-band images of the ESO Imaging
Survey (EIS, see Renzini \& da Costa 1997), using the matched filter
algorithm of Postman et al. (1996). Very little is known on the
performance of this algorithm at $z \geq 0.5$, and many of the cluster
candidates may not be real. The spectroscopic redshifts and
confirmation of cluster candidates in the range $0.5 \leq z \leq 0.7$
is possible with 4m-class telescopes.

Here we report on preliminary results of new spectroscopic
observations of six EIS candidate clusters.  A complete description of
our survey and results will be published in Ramella et al. (in
preparation). The selected cluster candidates have estimated redshifts
(from the matched filter algorithm) $0.5 \leq z_{mf} \leq 0.7$.  We
observed these cluster fields with EFOSC2 at the 3.6~m ESO telescope
at La Silla, in Multi-Object Spectroscopy mode, during two nights in
February 1999, in average weather conditions and partial moonlight.

In total we determined redshifts for 67 galaxies, covering the range
$0.09 \leq z \leq 0.79$ (plus a redshift for a serendipitously found
QSO at $z=3.2$), with an average $\overline{z}=0.38$. Magnitudes of
these galaxies span the range $17.0 \leq m_I \leq 21.3$, where $m_I$
is the apparent magnitude in the $I_c$ band (Nonino et al. 1999).

At the average estimated redshift of our candidate clusters, $z \sim
0.6$, the EFOSC2 field-of-view covers $1.9 \times 1.3$ h$_{75}^{-2}$
Mpc$^2$, roughly matching the typical size of clusters.  Therefore, in
searching for the redshift-space system that should correspond to the
2D EIS cluster, we consider the whole EFOSC2 field.

We start by defining as candidate galaxy systems, any set of two or
more galaxies in an EFOSC2 field, contained within a suitable redshift
range, $\Delta z$.  We use $\Delta z = 0.01 \times (1+z)$ (the $(1+z)$
factor is the usual cosmological correction -- see Danese et
al. 1980). Then, we estimate the likelihoods of the detected systems.
We compare the observed number of galaxies within each system against
the number of system galaxies expected for a uniform galaxy
distribution within our magnitude range. The luminosity function we
use is that of Postman et al. (1996), which, for our purposes, should
be close enough to the luminosity function of the EIS survey.  Since
field galaxies are inhomogeneously distributed, we calibrate the
likelihoods of our systems by comparison with a real field galaxy
sample (the Canada-France Redshift Survey, Lilly et al. 1995).

We find 4 real systems (at the 94~\% confidence level) among the six
EIS candidate clusters.  Note that the non-detection of two
of the candidate clusters does not prove the clusters do not exist. We
may simply have not been observing deep enough.  For the confirmed
clusters, only in two cases the spectroscopic mean redshift,
$\overline{z}$ is in agreement with the matched-filter estimate
($\overline{z}=0.673$ vs.  $z_{mf}=0.6$, and $\overline{z}=0.445$
vs. $z_{mf}=0.5$). In the other two, the spectroscopic redshift is
significantly smaller ($\overline{z}=0.129$ vs.  $z_{mf}=0.5$, and
$\overline{z}=0.236$ vs. $z_{mf}=0.5$).  Our spectroscopic results are
supported by independent evidence coming from the analysis of the
colour-magnitude diagrams of galaxies in these same cluster fields.

From our analysis we conclude that 2/3 of the candidate clusters we
have observed are real, and half of them have $\overline{z} \simeq
z_{mf}$.  Our sample is extremely small, but if we take our results at
face value, they imply that the EIS sample contains $\simeq 50$
clusters at $0.5 \leq \overline{z} \leq 0.7$.


\begin{references}
\reference Danese, L., De Zotti, G., di Tullio, G. 1980, \aap, 82, 322
%\reference Girardi, M., Biviano, A., Giuricin, G. et al. 1993, \apj, 404, 38
%\reference Holden, B.P., Nichol, R.C., Romer, A.K. et al. 1999, astro-ph/9907429
\reference Lilly, S.J., Le F\`evre, O., Crampton, D., Hammer, F., Tresse, L.
1995, \apj, 455, 50
\reference Nonino, M., Bertin, E., da Costa, L.N. et al. 1999, \aaps, 137, 51
\reference Olsen, L.F., Scodeggio, M., da Costa, L.N. et al. 1999, \aap, 345, 681
\reference Postman, M., Lubin, L.M., Gunn, J.E. et al. 1996, \aj, 111, 615
%\reference Ramella, M., Biviano, A., Boschin, W. et al. 2000, in preparation
\reference Renzini, A., da Costa, L.N. 1997, {\em Messenger,} 87, 23
\reference Scodeggio, M., Olsen, L.F., da Costa, L.N. et al. 1999, \aaps, 137, 83 
\end{references}
\end{document}